\documentclass{webofc}
\usepackage[varg]{txfonts}   

\woctitle{SQM2021} 

\def\lamb#1#2{$^{#1}_{\Lambda}${#2}}

\def\lamblamb#1#2{$^{~~#1}_{\Lambda\Lambda}${#2}}

\newcommand{\nopieft}{\mbox{$\slashed{\pi}$EFT}}

\usepackage{slashed}

\begin{document}

\title{Recent progress on few-body hypernuclei} 
\author{\firstname{Avraham} \lastname{Gal}\inst{1}\fnsep
\thanks{avragal@savion.huji.ac.il, invited talk at the 19th Int'l 
Conf. on Strangeness in Quark Matter (SQM2021), Brookhaven National 
Laboratory, Upton, New York, USA (online), May 2021: EPJ Web of Conferences 
{\bf 259}, 08002 (2022).} 
\institute{Racah Institute of Physics, the Hebrew University, Jerusalem
91904, Israel}} 

\abstract
{Few-body $\Lambda$ hypernuclei provide valuable information towards 
understanding strange matter. Recent experimental progress by the STAR 
Collaboration at the RHIC facility and by the ALICE Collaboration at the 
LHC has been matched by theoretical progress reviewed here: (i) lifetimes 
of the hypertriton \lamb{3}{H}, \lamb{3}{n} if particle-stable, \lamb{4}{H} 
and \lamb{4}{He} and their charge symmetry breaking, and (ii) the onset of 
$\Lambda\Lambda$ hypernuclear binding.} 
\maketitle

\section{Introduction} 
\label{sec:intro} 

Single- and double-$\Lambda$ hypernuclei provide unique extension of 
nuclear physics into strange hadronic matter~\cite{ghm16}. Experimental 
data on $\Lambda$ and $\Lambda\Lambda$ hypernuclei are unfortunately poorer 
both in quantity and quality than the data available on normal nuclei. 
Nevertheless, the few dozen $\Lambda$ separation energies $B_\Lambda$ of 
single-$\Lambda$ hypernuclei ($_{\Lambda}^{\rm A}$Z) determined across the 
periodic table from $A$=3 to 208, and the three $\Lambda\Lambda$ hypernuclei 
($^{~\rm A}_{\Lambda\Lambda}$Z) firmly established so far~\cite{GM11}, provide 
useful testground for the role of strangeness in dense hadronic matter, say in 
neutron star matter. Particularly meaningful tests of hyperon-nucleon and 
hyperon-hyperon strong-interaction models are possible in light $\Lambda$ 
and $\Lambda\Lambda$ hypernuclei, $A\leq 6$, which the topics reviewed 
below are concerned with.

\section{\lamb{3}{H} lifetime} 
\label{sec:l3h} 

Measurements of the \lamb{3}{H} lifetime in emulsion or bubble-chamber
experiments during the 1960s and early 1970s gave conflicting and puzzling
results. Particularly troubling appeared a conference report by Block
{\it et al.} claiming a lifetime of $\tau$(\lamb{3}{H})=(95$^{+19}_{-15}
$)~ps~\cite{Block62}, to be compared with a free $\Lambda$ lifetime $\tau_{
\Lambda}$=(236$\pm$6)~ps~\cite{Block63} measured in the same He bubble chamber
(BC). However, the He BC experiment~\cite{Keyes73} concluding that era, 
coauthored by the same Block, reported a value of $\tau$(\lamb{3}{H})=(246$
^{+62}_{-41}$)~ps, in agreement with a $\Lambda$ lifetime of (263$\pm$2)~ps. 
Given a weakly bound $\Lambda$, $B_\Lambda$(\lamb{3}{H})=0.13$\pm$0.05~MeV 
from emulsion studies~\cite{Davis05}, it was anticipated that 
$\tau$(\lamb{3}{H})$\approx$$\tau_\Lambda$. Recent measurements of $\tau
$(\lamb{3}{H}) in relativistic heavy ion collision experiments renewed 
interest in the \lamb{3}{H} lifetime problem. The first round of results 
from STAR~\cite{STAR10} and ALICE~\cite{ALICE16} suggested that $\tau
$(\lamb{3}{H}) is shorter than $\tau_{\Lambda}$ by as much as (30$\pm$17)\%. 
Whereas STAR's most recent published lifetime, from Au-Au collisions at 
$\sqrt{s_{\rm NN}}$=7.7 to 200~GeV, is even shorter than that: $\tau
$(\lamb{3}{H})=142$^{+24}_{-21}\pm$29~ps~\cite{STAR18}, ALICE latest published 
lifetime, from Pb-Pb collisions at $\sqrt{s_{\rm NN}}$=5.02~TeV, is close to 
$\tau_{\Lambda}$: $\tau$(\lamb{3}{H})=242$^{+34}_{-38}\pm$17~ps~\cite{ALICE19}, 
or 254$\pm$15$\pm$17~ps as reported in ICHEP 2020~\cite{Maz21}. The latest 
news is that this STAR-ALICE apparent divergence appears to be resolved by new 
STAR \lamb{3}{H} and \lamb{4}{H} lifetime measurements in Au-Au collisions at 
$\sqrt{s_{\rm NN}}$=3 and 7.2~GeV, resulting in $\tau$(\lamb{3}{H})=221$\pm
$15$\pm$19~ps~\cite{STAR21}. 

\begin{table}[hbt] 
\begin{center}
\caption{Calculated hypertriton weak decay rates 
$\Gamma$(\lamb{3}{H}$_{\rm g.s.}$) in units of the free-$\Lambda$ decay rate 
$\Gamma_{\Lambda}$, using \lamb{3}{H}(${\frac{1}{2}^+}$) wavefunctions that 
satisfy $B_\Lambda$(\lamb{3}{H})=0.13$\pm$0.05~MeV from emulsion studies 
\cite{Davis05}, and two-body branching ratios R$_3$=$\Gamma$(\lamb{3}{H}$\to
\pi^-+^3$He)/$\Gamma$(\lamb{3}{H}$\to\pi^-$+all). The underlined R$_3$ value 
is the BC world-average experimental value~\cite{Keyes73}.} 
\begin{tabular}{ccccc}
\hline
 Source & Method & $\pi$FSI & R$_3$ & 
$\Gamma$($_{\Lambda}^3$H)/$\Gamma_{\Lambda}$  \\ 
\hline  
 RD (1966) \cite{RD66} & closure-$\Lambda$pn & no & -- & 1.14  \\ 
 Congleton (1992) \cite{Cong92} & closure-$\Lambda$d & no & 0.33$\pm$0.02 & 
1.15  \\
 Kamada (1998) \cite{Kamada98} & Faddeev-$\Lambda$pn & no & 0.379 & 1.06  \\ 
 GG (2019) \cite{GG19} & closure-Faddeev-$\Lambda$pn & no & 0.362 & 
1.11$\pm$0.01  \\ 
 HH (2020) \cite{HH20} & $\slashed{\pi}$EFT(LO)-$\Lambda$d & no & 
0.37$\pm$0.05 & 0.98$\pm$0.15 \\  
\hline 
 GG (2019) \cite{GG19} & closure-Faddeev-$\Lambda$pn & yes & 0.357 & 
1.23$\pm$0.02  \\ 
 POGFG (2020) \cite{POGFG20} & $\pi$EFT(LO) $\Lambda$pn+$\Sigma NN$ & yes & 
\underline{0.35$\pm$0.04} & 1.38$^{+0.18}_{-0.14}$  \\ 
\hline 
\end{tabular} 
\label{tab:l3h} 
\end{center} 
\end{table} 

A taste of what Theory has to say about the hypertriton lifetime is 
demonstrated by a representative selection of \lamb{3}{H}$_{\rm g.s.}
({\frac{1}{2}}^+)$ lifetime calculations assembled in Table~\ref{tab:l3h}. Its 
upper part lists works that disregard pion final-state interaction ($\pi$FSI). 
The closure approximation was used in Refs.~\cite{RD66,Cong92,GG19}, whereas 
the other two works~\cite{Kamada98,HH20} accounted microscopically for the 
outgoing nucleon phase space and FSI. The lifetimes $\tau$(\lamb{3}{H})=1/$
\Gamma$(\lamb{3}{H}) derived from this upper part are shorter than 
$\tau_{\Lambda}$ by less than 13\%. In contrast, the two works listed in 
the lower part of the table suggest \lamb{3}{H} lifetimes shorter than 
$\tau_{\Lambda}$ by more than 20\%, owing primarily to $\pi$FSI where 
outgoing-pion plane waves are superseded by realistic pion distorted 
waves. This enhances $\Gamma$(\lamb{3}{H}) by about 10\%~\cite{GG19}, 
or more realistically by 15\%~\cite{POGFG20}. Interestingly, the recent 
$\pi$EFT work~\cite{POGFG20} is the only one that bothered to address the 
small $\Sigma NN$ components, of order $\lesssim$1\% probability, in the 
$\Lambda NN$ dominated \lamb{3}{H}. A relatively large reduction of $\Gamma
$(\lamb{3}{H}) by $\approx$10\% was found, owing to interference between the 
$\Sigma\to N\pi$ and $\Lambda\to N\pi$ weak-decay amplitudes. This effect was 
disregarded by Kamada {\it et al.} \cite{Kamada98}, thereby rendering their 
widely cited lifetime questionable. 

\begin{figure}[!t] 
\begin{center} 
\includegraphics[width=0.5\textwidth,height=5.5cm]{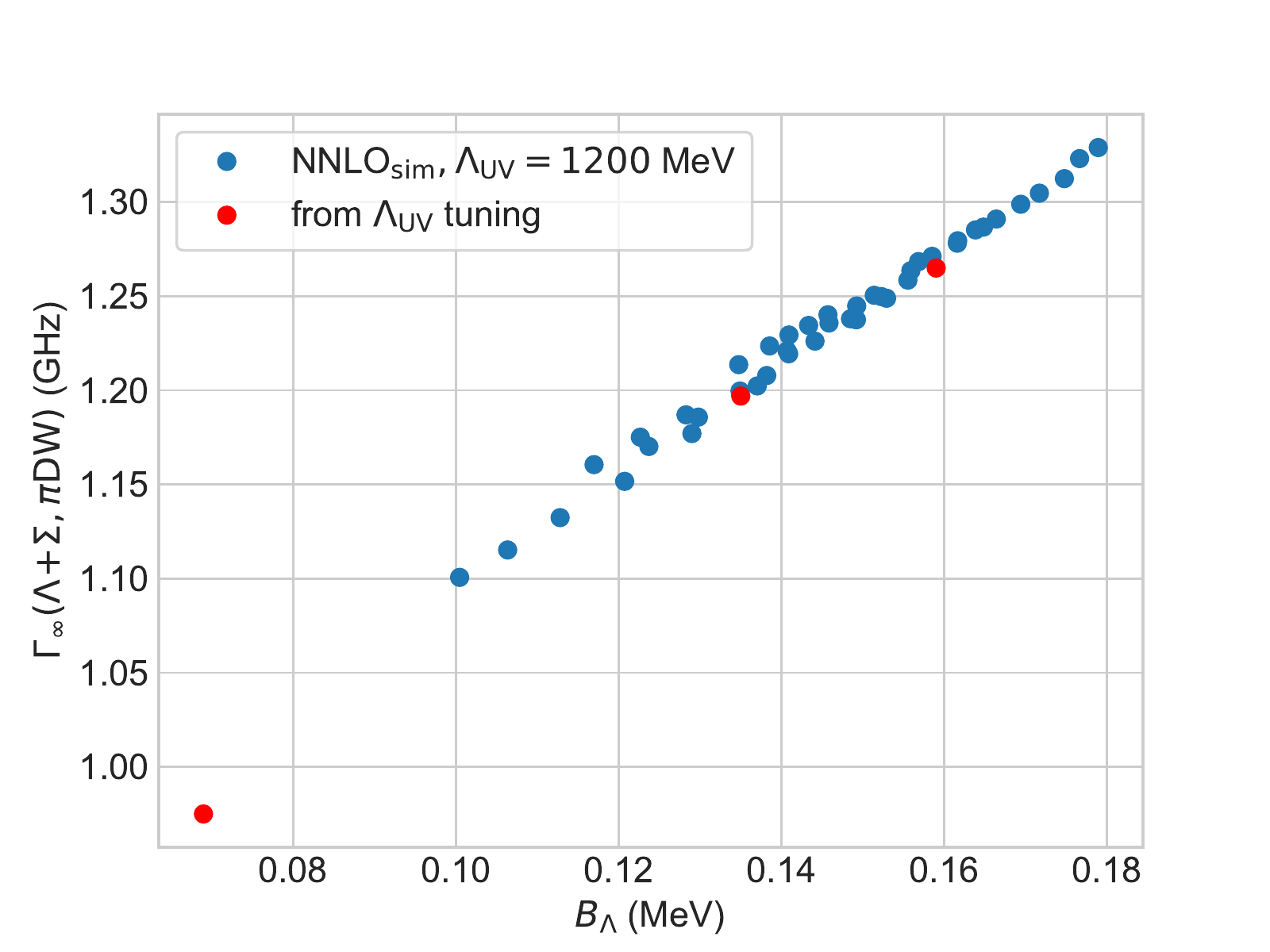} 
\caption{$\Gamma$(\lamb{3}{H}$\to\pi^-+^3$He) vs. $B_{\Lambda}$(\lamb{3}{H}) 
in $\pi$EFT. The three red points are from Ref.~\cite{POGFG20}, see text.} 
\label{fig:piEFT} 
\end{center} 
\end{figure}

The \lamb{3}{H} lifetime results shown in Table~\ref{tab:l3h} hold generally 
for \lamb{3}{H} wavefunctions corresponding to given $B_{\Lambda}$ values 
within the emulsion $B_\Lambda$(\lamb{3}{H})=0.13$\pm$0.05~MeV interval. 
The $B_{\Lambda}$ dependence of $\tau$(\lamb{3}{H}) was studied in two of 
these works, (i) with almost $B_{\Lambda}$-independent lifetime calculated 
in LO $\slashed{\pi}$EFT~\cite{HH20} and, in contrast, (ii) with rather 
strong $B_{\Lambda}$ dependence calculated in LO $\pi$EFT~\cite{POGFG20}, 
as shown in Fig.~\ref{fig:piEFT} provided by Daniel Gazda. The three red 
points in the figure correspond to three distinct values of $B_{\Lambda}
$(\lamb{3}{H}) reached by considering ultraviolet (UV) cutoff $\Lambda_{
\rm UV}$ values smaller than $\Lambda_{\rm UV}$=1200~MeV, beginning at 
which convergence is assured. This amounts to slightly varying the $\pi$EFT 
short-range input. The resulting $B_{\Lambda}$ dependence of the computed 
two-body decay rate $\Gamma$(\lamb{3}{H}$\to$$^3$He+$\pi^-$) is satisfied 
by the blue points, all for $\Lambda_{\rm UV}$=1200~MeV, obtained by varying 
systematically within allowed uncertainties some of the NN and YN chiral 
fit data. Of the three red points, the middle one is that highlighted 
in Table~\ref{tab:l3h} while the left one corresponds to $B_{\Lambda}
$(\lamb{3}{H})=0.069~MeV and $\tau$(\lamb{3}{H})=234$\pm$27~ps, a 
lifetime compatible perfectly with both ALICE and STAR new lifetime 
values \cite{Maz21,STAR21}. 

Given the strong correlation found in the $\pi$EFT calculation~\cite{POGFG20} 
between $B_{\Lambda}$(\lamb{3}{H}) and $\tau$(\lamb{3}{H}), a \lamb{3}{H} 
lifetime as close to $\tau_{\Lambda}$ supports ALICE preliminary value 
$B_{\Lambda}$(\lamb{3}{H})=0.05$\pm$0.06$\pm$0.10~MeV \cite{Caliva22} 
over STAR's published value 0.41$\pm$0.12$\pm$0.11~MeV \cite{STAR20}. 
As demonstrated in Ref.~\cite{POGFG20}, going to as high values of 
$B_{\Lambda}$(\lamb{3}{H}) would lead to considerably shorter values of 
$\tau$(\lamb{3}{H}), which are compatible perhaps with STAR's 2018 value 
\cite{STAR18}, but are incompatible with STAR's new value~\cite{STAR21}. 

\section{\lamb{3}{n} lifetime} 
\label{sec:l3n} 

\lamb{3}{n} was conjectured by the HypHI GSI Collaboration~\cite{Rap13} 
to be particle stable, while unstable unanimously in recent theoretical 
calculations~\cite{GV14,Hiyama14,GG14}. In \lamb{3}{n} decays induced by 
$\Lambda\to p+\pi^-$, where \lamb{3}{n} neutrons are spectators, the 
\lamb{3}{n}~$\to (pnn) + \pi^-$ weak decay rate is given in the closure 
approximation essentially by the $\Lambda\to p+\pi^-$ free-space weak-decay 
rate, whereas in $\Lambda\to n+\pi^0$ induced decays the production of a third 
low-momentum neutron is suppressed by the Pauli principle, so this \lamb{3}{n} 
weak decay branch may be disregarded up to perhaps a few percents. Hence 
$\Gamma({_\Lambda^3}{\rm n})/\Gamma_\Lambda \approx 1.114\times 0.641=
0.714$~\cite{GG19}, where the factor 1.114 follows from a difference between 
recoil energies in the \lamb{3}{n} and $\Lambda$ phase space factors, and the 
factor 0.641 is the free-space $\Lambda\to p+\pi^-$ fraction of the total 
$\Lambda\to N+\pi$ weak decay rate. This corresponds to a \lamb{3}{n} lifetime 
estimate of $\tau$(\lamb{3}{n})$\approx$368~ps, considerably longer than 
181${^{+30}_{-24}}\pm$25~ps or 190${^{+47}_{-35}}\pm$36~ps deduced from the 
$nd\pi^-$ and $t\pi^-$ alleged \lamb{3}{n} decay modes~\cite{Rap13,Saito16}, 
and thereby questioning the conjectured stability of \lamb{3}{n}.

\section{\lamb{4}{H} and \lamb{4}{He} lifetimes}
\label{sec:A=4}

A new STAR lifetime value for \lamb{4}{H} was reported in this meeting: 
$\tau$(\lamb{4}{H})=218$\pm$6$\pm$13~ps~\cite{STAR21}. Do we understand it in 
simple terms? A back-of-the-envelope estimate gives the following approximate 
expressions for \lamb{4}{H} (Z=1) and \lamb{4}{He} (Z=2) decay rates relative 
to the free $\Lambda$ decay: 
\begin{equation} 
\Gamma(^4_{\Lambda}{\rm Z})/\Gamma_{\Lambda}\approx (1+\eta(\bar q))\times 
(\alpha_{\rm Z}\times 0.7 + 1\times 0.3) + 0.20, 
\label{eq:l4Z} 
\end{equation}  
where $\eta(\bar q)\approx 0.50\pm 0.05$ is an exchange contribution to 
the dominant $s$-wave pionic weak decay rate, $\alpha_{\rm Z}=\frac{2}{3}$ 
($\frac{1}{3}$) for Z=1 (Z=2) by applying the $\Delta I$=$\frac{1}{2}$ rule 
to the $\pi ^4$He two-body decay modes, R$_4$=0.7 is the pionic two-body 
decay branching ratio~\cite{Bert70}, with 1$-$R$_4$=0.3 standing for the 
pionic multi-body part. Additional factors arising from recoil kinematics 
enhancement and $p$-wave decay suppression largely cancel out. Finally, 
the factor 0.20 stands for the observed $\Lambda N\to NN$ non-mesonic (n.m.) 
decay fraction $\Gamma_{\rm n.m.}/\Gamma_{\Lambda}$~\cite{Outa98,Parker07}. 
The resulting $\tau$(\lamb{4}{Z})=1/$\Gamma$(\lamb{4}{Z}) lifetime estimates 
\begin{equation} 
\tau_{\rm th}(^4_{\Lambda}{\rm H})=195\pm 10~{\rm ps}, \,\,\,\,\,\, 
\tau_{\rm th}(^4_{\Lambda}{\rm He})=263\pm 13~{\rm ps}, 
\label{eq:4th} 
\end{equation}
with a 5\% assigned theoretical uncertainty, are in good agreement with 
$\tau_{\rm exp}$(\lamb{4}{H})=194$\pm$25~ps and with $\tau_{\rm exp}
$(\lamb{4}{He})=256$\pm$27~ps as measured at KEK~\cite{Outa98} 
and confirmed, for \lamb{4}{He}, at BNL~\cite{Parker07}. A preliminary 
value $\tau_{\rm exp}$(\lamb{4}{H})=190$\pm$8~ps (stat. only) was derived 
in a test run of E73 at J-PARC~\cite{E73}.

\section{Charge symmetry breaking} 
\label{sec:csb}

\begin{figure}[!h] 
\begin{center} 
\includegraphics[width=0.5\textwidth,height=5cm]{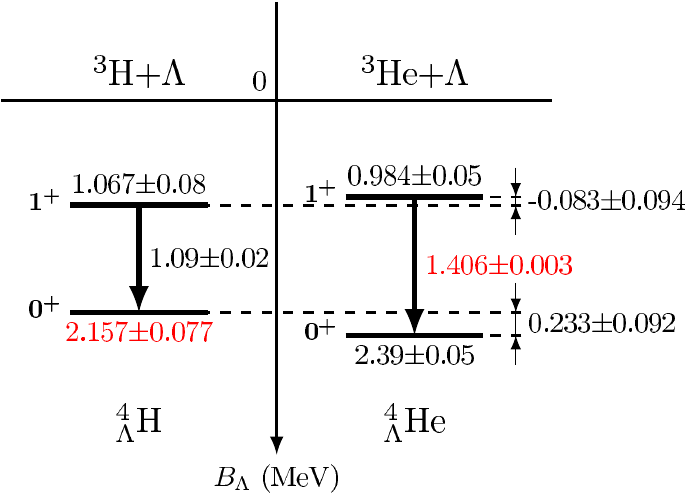} 
\caption{$A=4$ hypernuclear level scheme. Recent measurements of 
\lamb{4}{He}(1$^+$) excitation energy~\cite{E13} and of \lamb{4}{H}(0$^+_{\rm 
g.s.}$) binding energy~\cite{MAMI16} are marked in red. CSB splittings are 
shown to the right of the \lamb{4}{He} levels. Preliminary values from 
STAR are $\Delta B_{\Lambda}^{J=0}$=0.13$\pm$0.13$\pm$0.07~MeV, $\Delta 
B_{\Lambda}^{J=1}$=$-$0.19$\pm$0.13$\pm$0.07~MeV \cite{Shao21}.}
\label{fig:A=4}
\end{center}
\end{figure}

A special feature of the \lamb{4}{H}--\lamb{4}{He} mirror hypernuclei is 
the particularly strong charge symmetry breaking (CSB) reflected in their 
spectra, as shown in Fig.~\ref{fig:A=4}. As pointed out by Dalitz and von 
Hippel (DvH) one-pion exchange (OPE) contributes to the $\Lambda N$ potential 
only through a CSB component $V_{\rm CSB}^{\rm OPE}$ generated by admixing 
the SU(3) octet $\Lambda_{I=0}$ and $\Sigma^0_{I=1}$ hyperons in the physical 
$\Lambda$ hyperon~\cite{DvH64}. For the mirror \lamb{4}{H}--\lamb{4}{He} 
$0^+$, $1^+$ levels built on the mirror $^3$H-$^3$He g.s. cores, and using 
$A$=4 wavefunctions generated within a LO $\pi$EFT no-core shell-model 
calculation~\cite{gg16}, the regularized Yukawa tail of $V_{\rm CSB}^{
\rm OPE}$ gives rise to 
\begin{equation} 
{\rm OPE~(DvH):}\,\,\,\,\,\,\Delta B^{J=0}_{\Lambda}\approx 
175\pm 40~{\rm keV},\,\,\,\,\,\, \Delta B^{J=1}_{\Lambda}\approx 
-50\pm 10~{\rm keV},  
\label{eq:OPE} 
\end{equation}
where $\Delta B_{\Lambda}^J$$\equiv$$ B_{\Lambda}^J$(\lamb{4}{He})$-$$
B_{\Lambda}^J$(\lamb{4}{H}). Remarkably, the large $\Delta B^{J=0}_{\Lambda}$ 
OPE CSB (central plus tensor) contribution to the splitting of the 
\lamb{4}{H}--\lamb{4}{He} mirror g.s. levels roughly agrees with the observed 
value $\Delta B^{J=0}_{\Lambda}$=233$\pm$92~keV shown in Fig.~\ref{fig:A=4} 
which is considerably larger than the $\approx$70~keV CSB part of the 
Coulomb-dominated $\Delta B(^3$H--$^3$He)=764~keV in the mirror core nuclei, 
driven apparently by short-range $\rho^0$-$\omega$ mixing. Much smaller 
hypernuclear CSB contributions were found in $\Lambda NNN$ calculations 
by Coon {\it et al.}~\cite{Coon99}: $\Delta B^{J=0}_{\Lambda}
(\pi^0\eta+\rho^0\omega)\approx -$20~keV, $\Delta B^{J=1}_{\Lambda}
(\pi^0\eta+\rho^0\omega)\approx -$10~keV. Recent post-SQM21 EFT 
works \cite{HMN21,SBG22} representing CSB exclusively through two 
CSB low-energy constants are not discussed here. 

\section{Onset of $\Lambda\Lambda$ hypernuclear binding} 
\label{sec:LL} 

Reliable data on $\Lambda\Lambda$ hypernuclei are scarce. the Nagara emulsion 
event \cite{nagara01,ahn13} identified unambiguously as \lamblamb{6}{He}, 
with $\Delta B_{\Lambda\Lambda}$(\lamblamb{6}{He})=$B_{\Lambda\Lambda}
$(\lamblamb{6}{He})$-$2$B_{\Lambda}$(\lamb{5}{He})=0.67$\pm$0.17~MeV 
\cite{ahn13}, is the lightest particle-stable $\Lambda\Lambda$ hypernucleus 
found so far. Are there lighter particle-stable $\Lambda\Lambda$ species? 
This question has been addressed recently in a LO {\nopieft}~calculation 
\cite{csbgm19}, sketched below, and in a full coupled-channel $\pi$EFT 
calculation at NLO~\cite{LHMN21}, both concluding that the onset of 
$\Lambda\Lambda$ hypernuclear binding is given by the isodoublet 
\lamblamb{5}{H}--\lamblamb{5}{He}. 

\begin{figure}[htb]
\begin{center}
\includegraphics[width=0.48\textwidth,height=6cm]{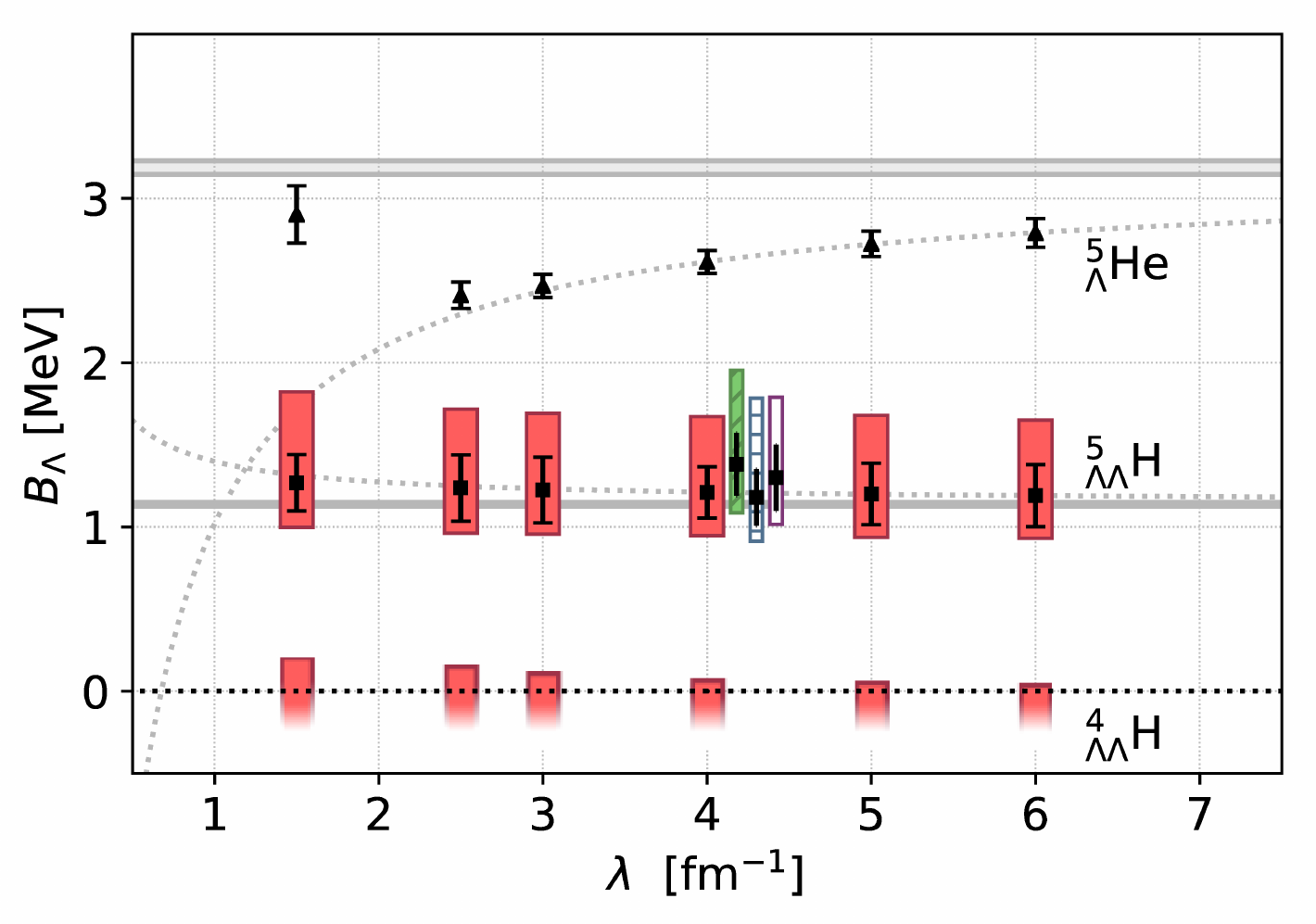}
\includegraphics[width=0.48\textwidth]{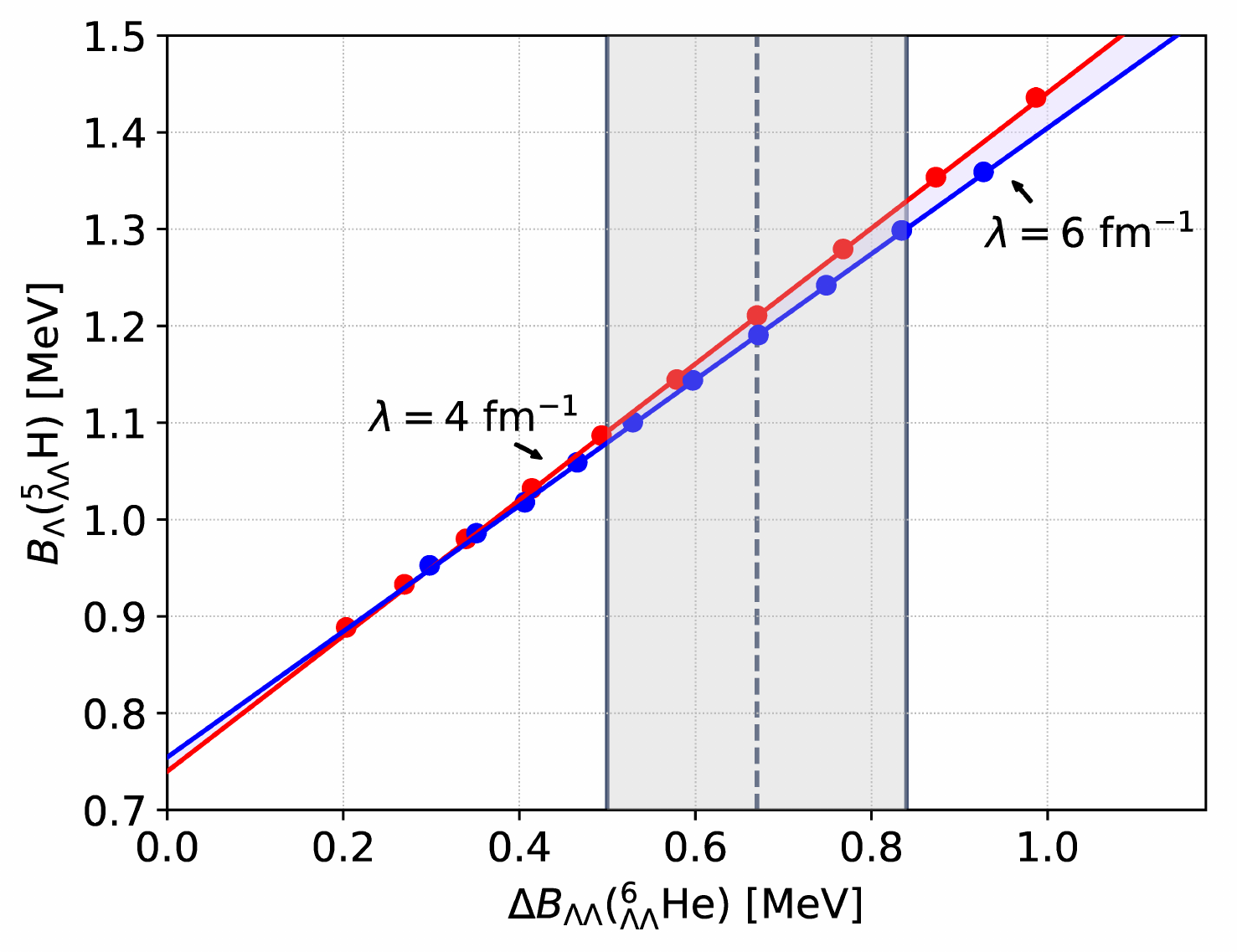}
\caption{Left: $B_{\Lambda}$(\lamb{5}{He},~\lamblamb{4}{H},~\lamblamb{5}{H})
from \nopieft~calculations~\cite{csbgm19}. Black error bars reflect given 
uncertainties in $^3_\Lambda$H, $^4_\Lambda$H, $^4_\Lambda$H$^\ast$, 
$^{~~6}_{\Lambda\Lambda}$He $B_{\Lambda}$ input data, red rectangles arise by 
varying $a_{\Lambda\Lambda}$ between $-$0.5 to $-$1.9~fm. Thin dotted lines 
show extrapolations to $\lambda\to\infty$ limits marked by gray horizontal 
bands. Right: Tjon lines calculations of $B_{\Lambda}$(\lamblamb{5}{H}) vs. 
$\Delta B_{\Lambda\Lambda}$(\lamblamb{6}{He}), with vertical straight lines 
marking the experimental uncertainty of $\Delta B_{\Lambda\Lambda}
$(\lamblamb{6}{He}).} 
\label{fig:LL5H}
\end{center}
\end{figure}

$\Lambda$ separation energy values $B_\Lambda(^{~~5}_{\Lambda\Lambda}$H) 
from \nopieft~calculations~\cite{csbgm19} are shown in the left panel of 
Fig.~\ref{fig:LL5H}. Several representative values of the $\Lambda\Lambda$ 
scattering length were used from a broad range of values suggested by analyses 
of $\Lambda\Lambda$ correlations derived recently in relativistic heavy-ion 
collisions and by analyzing the KEK-PS E522~\cite{Yoon07} invariant mass 
spectrum in the reaction $^{12}$C($K^-,K^+)\Lambda\Lambda X$ near the $\Lambda
\Lambda$ threshold; see Ref.~\cite{csbgm19} for references. Here the choice of 
$a_{\Lambda\Lambda}$ determines the one $\Lambda\Lambda$ low-energy constant 
(LEC) required at LO, while the $\Lambda\Lambda N$ LEC was fitted to the 
$\Delta B_{\Lambda\Lambda}(^{~~6}_{\Lambda\Lambda}$He)=0.67$\pm$0.17~MeV 
datum~\cite{ahn13}. The reported calculations used several $\Lambda N$ 
scattering-length combinations, demonstrating that the $\Lambda N$ model 
dependence is weak when it comes to double-$\Lambda$ hypernuclei, provided 
$B_\Lambda$ values of single-$\Lambda$ hypernuclei for $A<5$ are fitted to 
generate the necessary $\Lambda NN$ LECs. With values of $B_\Lambda(^{~~5}_{
\Lambda\Lambda}$H)$\sim$1~MeV, calculated over a broad range of cutoff values 
$\lambda$, it is clear that the particle stability of $^{~~5}_{\Lambda\Lambda}
$H is robust, in contrast to $^{~~4}_{\Lambda\Lambda} $H for which 
$B_\Lambda(^{~~4}_{\Lambda\Lambda}$H) comes out negative over most 
of the permissible parameter space in these calculations. Finally, 
Tjon-line correlations between $B_\Lambda(^{~~5}_{ \Lambda\Lambda}$H) 
and $B_\Lambda(^{~~6}_{\Lambda\Lambda}$He) are demonstrated in the 
right panel of Fig.~\ref{fig:LL5H}. We conclude that the isodoublet 
\lamblamb{5}{H}--\lamblamb{5}{He} marks the onset of $\Lambda\Lambda$ 
hypernuclear binding. J-PARC Experiment E75 will search for 
\lamblamb{5}{H}~\cite{Fujioka21}.

\section*{Acknowledgments} 

Discussions with my good colleagues coauthoring 
Refs.~\cite{GG19,POGFG20,gg16,csbgm19} are gratefully acknowledged. 
This work is part of a project funded by the EU Horizon 2020 Research 
\& Innovation Programme under grant agreement 824093.

\end{document}